\documentclass[twocolumn,showpacs,preprintnumbers,amsmath,amssymb,aps]{revtex4-1}

\usepackage{graphicx}
\usepackage{bm}
\usepackage{subfigure}
\usepackage{natbib}

\begin{document}


\title{$\alpha$-Scale Decoupling of the Mechanical Relaxation and Diverging Shear Wave Propagation Lengthscale in Triphenylphosphite}

\author{Darius H. Torchinsky}
\affiliation{Department of Physics, Massachusetts Institute of
Technology, Cambridge, Massachusetts 02139}
\author{Jeremy A. Johnson}
\affiliation{Department of Chemistry, Massachusetts Institute of
Technology, Cambridge, Massachusetts 02139}
\author{Keith A. Nelson}
\email{kanelson@mit.edu} \affiliation{Department of Chemistry,
Massachusetts Institute of Technology, Cambridge, Massachusetts
02139}

\date{\today}

\begin{abstract}
We have performed depolarized Impulsive Stimulated Scattering
experiments to observe shear acoustic phonons in supercooled
triphenylphosphite (TPP) from $\sim$10 - 500~MHz. These
measurements, in tandem with previously performed longitudinal and
shear measurements, permit further analyses of the relaxation
dynamics of TPP within the framework of the mode coupling theory
(MCT). Our results provide evidence of $\alpha$ coupling between
the shear and longitudinal degrees of freedom up to a decoupling
temperature $T_c = 231~K$. A lower bound length scale of shear
wave propagation in liquids verified the exponent predicted by
theory in the vicinity of the decoupling temperature.


\end{abstract}


\maketitle

\section{\label{sec:intro}Introduction}

The most prominent characteristic of the frequency relaxation
spectrum of glass forming liquids is the $\alpha$ peak, a broad,
yet distinct feature whose origin lies in the collective degrees
of freedom which undergo kinetic arrest during vitrification
\cite{binderkob,donth,angellreview}. As the transition from liquid
to glass occurs either upon cooling or the application of high
pressure, the characteristic time scale of $\alpha$ relaxation
varies by 16 orders of magnitude, from the $\sim$100~fs timescale
associated with an attempt frequency in the liquid state
\cite{uli} to the 100~s timescale (arbitrarily) chosen at $T_g$
\cite{angellreview}. One of the fundamental tasks presented to the
spectroscopy of glass forming liquids is thus the complete
characterization of the $\alpha$ relaxation across this span of
frequencies.

To date, this challenge has best been met by dielectric
spectroscopy, whose wide dynamical range remains unparalleled in
the study of glass forming liquids
\cite{lunkenheimer,loidl1,dielmct}. Nevertheless, dielectric
spectroscopy of molecular liquids probes mainly orientational
relaxation, while the glass transition is more naturally
understood in terms of the degrees of freedom associated with
flow, i.e. the density, which may be probed by longitudinal
acoustic waves, and the transverse current, which couples to shear
waves \cite{balucani,boonyip}.

Despite the importance of both of these degrees of freedom, the
bulk of the literature on structural relaxation has focussed on
study of the longitudinal modulus due to the relatively limited
means by which shear waves may be generated and probed. Acoustic
transducer \cite{ferry,read}, Impulsive Stimulated Scattering
(ISS) \cite{nelson,yan:6240,yang1} and high-frequency forced
Brillouin Scattering \cite{deathstar} techniques can continuously
cover the longitudinal spectrum up to the mid-GHz regime. However,
broadband shear wave generation is more accessible experimentally
at lower frequencies and exhibits a gap in the range from the
$\sim 10$~MHz limit of transducer methods \cite{ferry,read} to the
low GHz frequencies probed by depolarized (VH) Brillouin
Scattering \cite{huang,drey1,drey2,SAL2,wang:617,tao}. The vast
majority of depolarized Brillouin Scattering studies are conducted
in a single scattering geometry, permitting observation of only
one acoustic wavevector. Thus, relaxation information must be
extracted by complex modeling incorporating contributions from the
various channels responsible for the depolarized scattering of
light. Obtaining spectral information from a collection of
acoustic frequencies would allow a direct determination of the
shear relaxation spectrum without the need for multiparameter
fitting.

Recent advancements in the generation of narrowband shear waves
\cite{torchinsky1} using the technique of Impulsive Stimulated
Brillouin Scattering (ISBS) in the depolarized geometry
\cite{nelson2} now permit study of the shear $\alpha$ relaxation
spectrum across a broad range of frequencies. In this study, we
employ ISBS to focus on tests of a first-principles theory of the
glass transition, the mode coupling theory (MCT)
\cite{bengt,goetze1988,ab}. While the natural variables of the
theory are the density fluctuations that constitute the
longitudinal modes, MCT suggests relationships among the
temperature dependencies of the various relaxing variables (e.g.,
transverse current fluctuations, density fluctuations,
orientational fluctuations, etc.) through the property of
$\alpha$-scale coupling \cite{ab}. This prediction states that if
two variables, $A$ and $B$, couple to density fluctuations, their
associated relaxation times, $\tau_{A}(T)=C_A \tau(T)$ and
$\tau_B(T)=C_B \tau(T)$, are equivalent, up to a temperature
independent factor, to a universal relaxation time $\tau(T)$
characteristic of the main $\alpha$ relaxation process. Decoupling
may then occur at the MCT critical temperature $T_c$ when the
relationship between these relaxation times can break down. Prior
experimental studies that have examined $\alpha$-coupling have
focused on, e.g., the coupling of dielectric to rheological
variables \cite{zorn}, dielectric to shear mechanical relaxation
\cite{niss2}, and of translational to rotational diffusion
\cite{lohfink:30}. However, there have been no studies that
directly link elastic degrees of freedom with each other.

MCT predictions have also been formulated concerning the
transverse current fluctuations in their own right, in particular
the power-law divergence of an upper bound length scale for shear
wave propagation derived for hard spheres in a Percus-Yevick
approximation \cite{das1}. While such upper bound has been
established in the literature \cite{balucani} in terms of a
wavelength at which the corresponding shear acoustic frequency
goes to zero, this particular scaling relationship with
temperature is unique to MCT and, to our knowledge, has not
undergone scrutiny in the lab.

Below, we provide a test of these assertions by building upon
prior work \cite{sil1,sil2}, and present an expanded study of the
shear acoustic behavior of the fragile glass-former
triphenylphosphite (TPP) ($T_g$~=~202~K \cite{TPPDI}). Taken
together, these experiments fill in the gap between 10~MHz and
1~GHz in the shear relaxaiton spectrum, and thus constitute the
broadest bandwidth shear acoustic measurements performed optically
any glass forming system to our knowledge. Our measurement
approach thus enables characterization of shear relaxation in
glass-forming liquids in this frequency regime. Although
supercooled liquid $\alpha$ relaxation spectra are typically
broader than two decades at any temperature and, as $T$ is varied,
the spectra move across many decades, the 10-1000~MHz range is
sufficient to permit assessment of whether shear relaxation
dynamics are consistent with relaxation dynamics determined from
longitudinal wave measurements covering a comparable spectral
width at common temperatures. They are also sufficient to test
predictions about the low-frequency limit of shear wave
propagation as a function of temperature.

We begin with a general overview of depolarized ISBS. This
motivates a derivation of the signal generated in an ISBS
experiment geared toward measuring shear waves from simplified
versions of the hydrodynamic equations of motion. After a brief
review of the experimental apparatus used, we present data
collected in the temperature range from 220~K to 250~K. Our
measured shear acoustic frequencies show significant acoustic
dispersion, indicative of the complex relaxation dynamics
characteristic of supercooled liquids thus providing a means of
testing the MCT predictions described above. In particular, we see
evidence of $\alpha$-decoupling, yielding an estimate of the MCT
$T_c$ of 231~K. Using this value of the crossover temperature
yields a power law for the lower limit of shear wave propagation
in line with predictions derived in Ref.~\cite{das1}. These
results are discussed within the framework of the MCT.

\section{\label{sec:gen}ISBS -- General Considerations}

\begin{figure}
\centering
\includegraphics[scale=0.32]{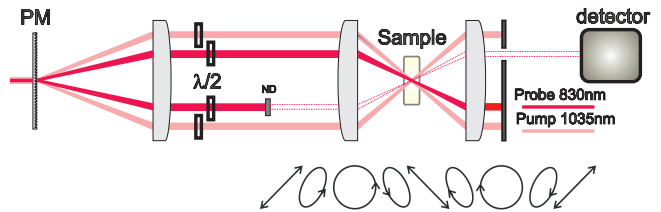}
\caption[ISBS setup]{\label{fig:expt}Schematic illustration of the
ISS setup. Both the pump and probe beams are incident on the PM
and their $\pm 1$ diffraction orders are recombined at the sample.
In the case of a depolarized experiment, half-waveplates in the
path of each of the four beams are used to create a polarization
grating pattern in the sample plane, as shown in the figure
inset.}
\end{figure}

In a typical ISS experiment, conducted in a heterodyned four-wave
mixing geometry, light from a pulsed laser is incident on a
diffractive optical element, typically a binary phase mask (PM)
pattern, and split into two parts ($\pm 1$ diffraction orders;
other orders are blocked) that are recombined at an angle $\theta$
as depicted in Fig.~\ref{fig:expt}. The crossed excitation pulses
generate an acoustic wave whose wavelength $\Lambda$ is given by
\begin{equation}
\Lambda=\frac{\lambda}{2\sin{\theta/2}}
\end{equation}
where $\lambda$ is the excitation laser wavelength. Probe light
(in the present case from a CW diode laser) is also incident on a
phase mask pattern (the same one or another with the same spatial
period) and split into two parts that are recombined at the sample
to serve as probe and reference beams. The signal arises from
diffraction of probe light off the acoustic wave and any other
spatially periodic responses induced by the excitation pulses. The
diffracted signal field is superposed with the reference field for
heterodyned time-resolved detection of the signal, which typically
shows damped acoustic oscillations from which the acoustic
frequency and damping rate can be determined.

The polarizations of the beams in the present measurements are all
vertical (V) or horizontal (H) relative to the scattering plane.
We denote the polarizations of the excitation fields, probe field,
and reference/signal field in that order, i.e. VHVH denotes V and
H excitation, V probe, and H reference and signal polarizations
while VVVV denotes all V polarizations. All the measurements
reported herein were conducted in the VHVH configuration.

In a depolarized (VHVH) experiment, each of the pump arms carries
a different polarization and the ensuing grating is described as
an alternating polarization pattern, depicted in the
Fig.~\ref{fig:expt} inset. It is the regions of linear
polarization that perform electrostrictive work, deforming the
excited region in a fashion that generates counterpropagating
shear acoustic waves with a driving force that scales with
wavevector magnitude $q=2\pi/\Lambda$ \cite{nelson2}. The driving
force also scales with a geometrical factor of $\cos{\theta/2}$
due to a diminishing of the contrast ratio of the polarization
grating with increased angle. In the small $\theta$ limit, the
signal from shear waves becomes markedly stronger as the
wavevector is increased. In this case, the diffracted signal
polarization is rotated $90^{\circ}$ from that of the incident
probe light, analogous to depolarized (VH) Brillouin scattering.
Finally, we note that the excitation pulses may also induce
molecular orientational responses that can contribute to signal,
analogous to depolarized quasielastic scattering
\cite{sil1,fayer,pick1,pick2,pick3,pick4,azzimani}.

\section{\label{sec:thry}ISBS -- Theoretical Background}

A full treatment of signal in an ISS experiment includes
longitudinal and transverse acoustic modes, molecular
orientations, and other nonoscillatory degrees of freedom
\cite{yang1,pick1,pick2,pick3,pick4,azzimani,glorieuxflow}. Here,
we present a slightly simplified derivation that allows us to
understand the presence of shear and orientational contributions
to the signals.

We start with the generalized, linearized hydrodynamic equations
of motion \cite{hansmc}. First is the momentum density
conservation law,
\begin{equation}
\frac{\partial}{\partial t}\rho \textbf{u}(\textbf{r},t) + \nabla
\cdot \Pi (\textbf{r},t) = 0
\end{equation}
where $\rho$ is the density, $\textbf{u}(\textbf{r},t)$ the
velocity, and $\Pi (\textbf{r},t)$ the stress tensor. The
expression for the stress tensor reads
\begin{eqnarray}\label{eq:bem}
\Pi^{\alpha \beta} (\textbf{r},t)&=&\delta_{\alpha
\beta}P(\textbf{r},t) - \eta \tau_{\alpha
\beta}(\textbf{r},t)\nonumber\\
&&+\delta_{\alpha \beta}(\frac{2}{3}\eta
-\zeta)\textbf{$\nabla\cdot$ u }(\textbf{r},t)\nonumber\\
&-&\mu Q_{\alpha \beta}(\textbf{r},t)+F_{\alpha
\beta}(\textbf{r},t).
\end{eqnarray}
Here, $\delta_{\alpha \beta}$ is the Kronecker delta,
$P(\textbf{r},t)$ is the pressure, $\eta$ is the shear viscosity,
$\zeta$ is the longitudinal viscosity, $Q_{\alpha
\beta}(\textbf{r},t)$ is the orientational variable, $\mu$
expresses the coupling of translational force due to rotational
motion, and $\textbf{r}$ is the spatial coordinate. $F_{\alpha
\beta}(\textbf{r},t)$ is the external shearing stress of the laser
field \cite{ferry,wang}, assumed to be temporally impulsive and
spatially periodic, i.e.,
$F_{\alpha\beta}=F_0^{xy}\delta(t)\cos{\textbf{q}\cdot\textbf{r}}$,
and $\tau_{\alpha \beta}$ is the rate of strain, defined by
\begin{equation}
\tau_{\alpha \beta}=\frac{\partial u_\alpha
(\textbf{r},t)}{\partial r_\beta}+\frac{\partial u_\beta
(\textbf{r},t)}{\partial r_\alpha}.
\end{equation}

The orientational variable obeys its own equation of motion, given
by
\begin{equation}\label{eq:qem}
\frac{\partial Q_{\alpha \beta}(\textbf{r},t)}{\partial t} =
-\Gamma_R Q_{\alpha \beta}(\textbf{r},t)+\xi \tau_{\alpha
\beta}(\textbf{r},t)+Q_0(\textbf{r},t)
\end{equation}
where $\Gamma_R$ is the orientational relaxation rate, $\xi $ is
the torque due to translational motion, and $Q_0(\textbf{r},t)$ is
the torque exerted by the laser \cite{wang,wang:617}. As with the
laser-induced stress, the torque will also be modeled as
temporally impulsive and spatially periodic. This equation of
motion assumes Debye orientational relaxation as a computational
convenience.

We set the grating wavevector in the $x$ direction and the
transverse direction to be $y$. Since we are interested in shear
waves, we select the transverse elements of the above equations,
and after a Fourier-Laplace transform defined as
\begin{equation}\label{eq:3fourlap}
{\cal
FL}\{f(\textbf{r},t)\}=\int_{0}^{\infty}dt\int_{-\infty}^{\infty}
d\textbf{r}f(\textbf{r},t)e^{i\textbf{q}\cdot\textbf{r}-st},
\end{equation}
we are left with
\begin{eqnarray}
\rho s u_y(\textbf{q},s) &=&-i q \Pi^{xy}(\textbf{q},s)\\
\Pi^{xy}(\textbf{q},s) &=&-\eta_s \tau^{xy}(\textbf{q},s) + F_0^{xy}+\mu Q_{xy}(\textbf{q},s) \\
s Q_{xy}(\textbf{q},s)&=&-\Gamma_R Q_{xy}(\textbf{q},s) +\xi
\tau^{xy}(\textbf{q},s)+Q_0.
\end{eqnarray}

For simplicity, we assume that the measured signal is proportional
to the molecular polarizability anisotropy \cite{wang}, i.e. to
$Q_{\alpha\beta}$, so we solve for the orientational variable to
yield
\begin{equation}
Q_{xy}(\textbf{q},s)=\frac{\xi q^2 F_0^{xy} +Q_0(s\rho
+q^2\eta)}{(s+\Gamma_R)(s\rho +q^2\eta)-\xi\mu q^2}.
\end{equation}
In order to arrive at an analytic solution, we make the
approximation that $\xi\mu q^2$ is a coupling of higher order that
can be ignored in the solution of the equations of motion.
Physically, we are making the approximation that we may ignore the
recoupling of the orientational degree of freedom to itself via
rotational-translational coupling. In this approximation, the
expression for $Q_{xy}$ separates as
\begin{equation}
Q_{xy}(\textbf{q},s) = \frac{\xi q^2F_0^{xy}}{(s+\Gamma_R)(s\rho
+q^2\eta)}+\frac{Q_0}{s+\Gamma_R}.
\end{equation}

In order to extract the effect of structural relaxation dynamics
on signal, we model the $\alpha$ peak by Debye relaxation as
$\frac{c_{\infty}^2 \tau_s}{1+s\tau_s}$, where $c_\infty$ is the
infinite frequency speed of sound and $\tau_s$ is the
characteristic shear relaxation time. This, in effect, makes
$\eta$ complex, which enables an elastic component of the shear
response to emerge from the equation of motion~\ref{eq:bem}
\cite{hansmc}. In this model, the above equations can be solved
for $Q_{xy}$ to yield
\begin{equation}\label{eq:eps}
Q_{xy}(\textbf{q},s)=\frac{\xi
q^2F_0^{xy}}{\rho}\frac{1}{(s+\Gamma_R)(s + \frac{c_\infty^2
\tau_s q^2}{1+s\tau_s})}+\frac{Q_0}{s+\Gamma_R}.
\end{equation}
Equation (\ref{eq:eps}) can be recast into the form:
\begin{eqnarray}\label{eq:eps2}
Q_{xy}(\textbf{q},s)&=&\frac{\xi q^2F_0^{xy}}{\rho}\frac{2
\Gamma_s +s}{(s+\Gamma_R)(s+\Gamma_s+i\omega_s)(s+\Gamma_s-i\omega_s)}\nonumber\\
&&+\frac{Q_0}{s+\Gamma_R}
\end{eqnarray}
where the shear acoustic damping rate $\Gamma_s$ is given by
\begin{equation}\label{eq:ga}
\Gamma_s=\frac{1}{2\tau_s}
\end{equation}
and the frequency of oscillation $\omega_s$ by
\begin{equation}\label{eq:wa}
\omega_s = \sqrt{c_\infty^2 q^2-\left(\frac{1}{2\tau_s}\right)^2}.
\end{equation}
The acoustic damping rate $\Gamma_s$ thus comprises the structural
relaxation dynamics. We also note that $\omega_s$ may go to zero
for finite wavevector $q_0$ when
\begin{equation}\label{eq:cicond}
q_0 = \frac{1}{2\tau_s c_{\infty}}.
\end{equation}

Separation of equation \ref{eq:eps2} by partial fractions yields a
time-domain solution
\begin{eqnarray}\label{eq:shearsol}
Q_{xy}(q,t)&=& \frac{\xi q^2 F_0^{xy}}{\rho}[Ae^{-\Gamma_s
t}\sin(\omega_s t)\nonumber\\
&&+B\left(e^{-\Gamma_R
t}-e^{-\Gamma_s t}\cos(\omega_s t)\right)]\nonumber\\
&&+Q_0\exp(-\Gamma_R t)
\end{eqnarray}
where
\begin{equation}
A=\frac{\omega_s^2+\Gamma_R\Gamma_s-\Gamma_s^2}{\omega_s\left(\left(\Gamma_s-\Gamma_R\right)^2+\omega_s^2\right)}
\end{equation}
and
\begin{equation}
B=\frac{(2\Gamma_s-\Gamma_R)}{\left(\left(\Gamma_s-\Gamma_R\right)^2+\omega_s^2\right)}.
\end{equation}

The solution represented by equation~\ref{eq:shearsol} comprises
two pieces. The term proportional to $q^2 F_0^{xy}$ is due to the
shear acoustic response, and the other term proportional $Q_0$ is
a decaying exponential independent of $q$. This orientational
response is the optical kerr effect (OKE) signal. We also note the
presence of an orientational contribution to signal in the
acoustic response, due to the rotational-translational coupling.

In order to derive the relaxation spectrum from our data, we
consider a frequency-dependent modulus $G^{\ast}(s)=G'(s)+iG''(s)$
which obeys the dispersion relation \cite{silence_thesis}
\begin{equation}\label{eq:gdisp3}
\rho s^2+G(s)q^2=0,
\end{equation}
and yields the following expressions for the real and imaginary
parts of the shear modulus from the acoustic frequency and damping
rate
\begin{eqnarray}\label{realmod}
G'(\omega_s)&=&\rho\frac{\omega_s^2-\Gamma_s^2}{q^2}\\
G''(\omega_s)&=&\rho\frac{2\omega_s\Gamma_s}{q^2}.\label{imagmod}
\end{eqnarray}
As equation \ref{eq:gdisp3} has been derived considering the
strain, in order to compare it with the results of the above
analysis, we must solve for the strain from the original equations
of motion. This gives the dispersion relation
\begin{equation}\label{eq:edisp3}
\rho s^2+\eta s q^2 =0.
\end{equation}
Comparison between equations \ref{eq:gdisp3} and \ref{eq:edisp3}
yields the connection between the elastic modulus and the
relaxation spectrum $\eta(q,\omega_s)$
\begin{eqnarray}
\frac{G'(\omega_s)}{\rho}&=&-\omega_s{\rm Im}[\eta(q,\omega_s)]\\
\frac{G''(\omega_s)}{\rho}&=&\omega_s{\rm Re}[\eta(q,\omega_s)].
\end{eqnarray}

We conclude by noting that orientational responses of anisotropic
molecules can be induced not only by the excitation pulses, as in
the case of OKE, but also by flow that occurs due to the induced
density changes \cite{glorieuxflow}. Both of these sources lead to
signals that can be suppressed by proper selection of probe and
signal polarizations. This step was impractical when measuring
shear responses, as a choice of polarization which would suppress
the orientational response often reduced the already weak shear
signal beyond the limit of detection. Eliminating this
contribution was also deemed unnecessary, as the aim of this study
was to directly probe the shear relaxation spectrum via narrowband
acoustic measurements of frequencies and damping rates.

\section{ISBS -- Experimental Details}

The pump laser system used for these studies was an Yb:KWG High-Q
FemtoRegen lasing at 1035~nm and producing pulses of 500~$\mu$J at
a repetition rate of 1~kHz, although 150~$\mu$J was routinely used
to avoid cumulative degradation of the sample. While a 300~fs
compressed pulse width FWHM was typical, we bypassed the
compressor to retrieve pulses directly from the regen that were
80~ps in duration in order to avoid sample damage at high peak
powers, yet remain in the impulsive limit relative to the
oscillation period. The excitations beams were cylindrically
focussed to a spot that was 2.5~mm in the grating wavevector
dimension and 100~$\mu$m in the perpendicular dimension so that
the acoustic waves would have many periods and the decay of signal
would be due primarily to acoustic damping rather than propagation
away from the excitation and probing region of the sample.

As a probe, we used a Sanyo DL8032-001 CW diode laser output at
830~nm with 150~mW power focused to a spot of 1~mm in the grating
dimension by 50~$\mu$m in the perpendicular dimension. We also
used a single phase mask optimized for diffraction into $\pm$1
order at 800~nm for both pump and probe beams, and we utilized
two-lens 2:1 imaging with Thorlabs' NIR achromats to recombine the
beams at the sample. The local oscillator was attenuated by a
factor of $10^{-3}$. Approximately $30\%$ of the pump power was
lost into zero order with this configuration, but the pump
intensity still had to be reduced significantly to avoid unwanted
nonlinear effects.

In order to generate the polarization grating, we inserted
$\lambda/2$ waveplates into each of the beams. These waveplates
were held in precision rotation mounts to provide accurate
alignment of the relative polarizations of the V and H polarized
beams. This set the upper limit on the grating spacing that could
be achieved in our measurements -- for longer wavelengths, the
beams came close enough together to be clipped by the rotation
mounts. This issue was addressed by imaging with longer focal
length optics. The signal was collected in a Cummings Electronics
Laboratories model 3031-0003 detector and recorded by a Tektronix
Model TDS-7404 oscilloscope. The shear signals were weak and
required 10,000 averages, resulting in data acquisition times of a
few minutes.

TPP at 97\% nominal purity was purchased from Alfa Aesar and had
both water and volatile impurities removed by heating under vacuum
with the drying agent ${\rm MgSO_4}$ immersed in the liquid. The
sample was then transferred to a cell with movable windows
\cite{halalay} via filtering through a millipore 0.22~$\mu$m
teflon filter. After loading, the cell was placed in a Janis
ST-100-H cryostat where the temperature was measured with a
Lakeshore model PT-102 platinum resistor immersed directly within
the liquid, and monitored and controlled with a Lakeshore 331
temperature controller.

The grating spacings examined in this study were 2.33~$\mu$m,
3.65~$\mu$m, 6.70~$\mu$m, 7.61~$\mu$m, 9.14~$\mu$m, 10.2~$\mu$m,
11.7~$\mu$m, 13.7~$\mu$m, 15.7~$\mu$m, 18.3~$\mu$m, 21.3~$\mu$m,
24.9~$\mu$m, 28.5~$\mu$m, 33.0~$\mu$m, 38.1~$\mu$m, 42.6~$\mu$m,
and 50.7~$\mu$m, while data for 0.48~$\mu$m, 1.52~$\mu$m,
3.14~$\mu$m, and 4.55~$\mu$m grating spacings were taken from
prior reported results \cite{sil1,sil2}. The acoustic wavelength
was calibrated through ISS measurements in ethylene glycol, for
which the speed of sound is known to a high degree of accuracy
\cite{silence_thesis}. When the samples were cooled to the desired
temperature, the cooling rate never exceeded 6~K/min, with 2~K/min
being typical. Data were taken at fixed wavevector every 2~K from
220~K to 250~K upon warming as we found crystallization was less
likely to occur upon warming than cooling. Only a few measurements
could be taken without having to thermally cycle the liquid, as it
invariably crystallized. We noticed that the tendency toward
crystallization was particularly pronounced in the temperature
range between 234~K and 242~K. After a few days of use, the sample
was observed to develop a slightly cloudy yellowish hue, and so
was replaced by a new one. The yellowish samples tended more
readily toward photoinduced damage, as well as crystallization,
than the original, clear samples. Comparison of the signals
obtained from the degraded samples and fresh ones yielded the same
frequency and damping rate values, indicating that uncertainties
in either of these quantities were due mainly to noise in the
data.

\section{\label{sec:randd}Results and Discussion}

\begin{figure}
\includegraphics[scale=0.82]{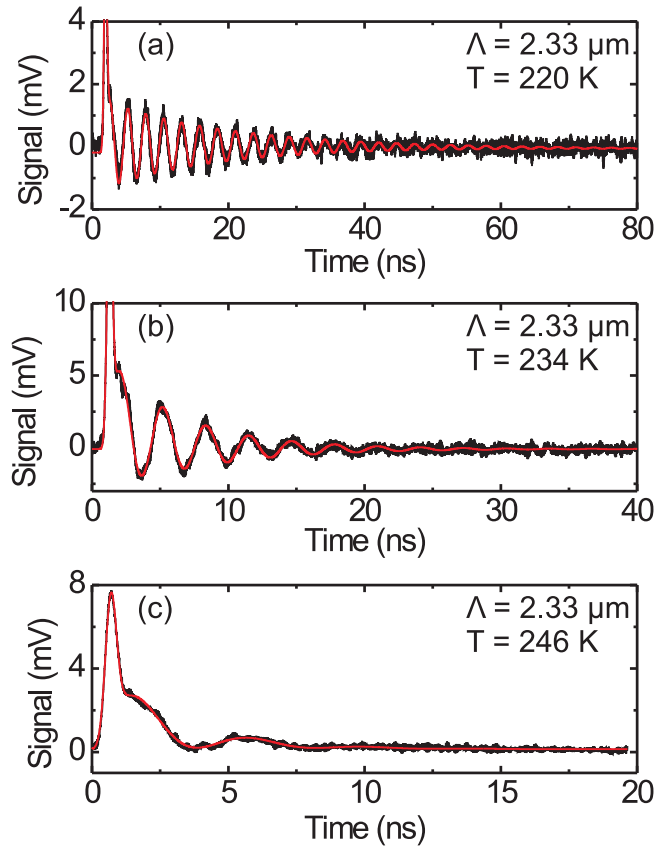}
\caption{\label{fig:TPP_2.4} Shear waves in TPP with $\Lambda
=2.33~\mu$m at (a) 220~K, (b) 234~K, and (c) 246~K. The data are
in black and the fits in red. As the temperature is increased, the
acoustic wave becomes more heavily damped. At higher temperatures,
we also note the presence of the orientational relaxation, which
appears to skew the signal such that the oscillations do not occur
about the zero baseline.}
\end{figure}

The results of several VHVH experiments performed at
$\Lambda$=2.33~$\mu$m grating spacing are shown in
Fig.~\ref{fig:TPP_2.4}. There is an initial spike due to the
non-resonant electronic response. Immediately following this
hyperpolarizability peak are oscillations about the zero baseline
from the counterpropagating shear waves. At a sample temperature
of 220~K, these oscillations are seen to disappear on the scale of
tens of nanoseconds due to acoustic damping. As the sample is
warmed, the frequency is observed to decrease and the damping to
increase dramatically. At sufficiently high temperatures, the
shear wave becomes overdamped. We also note that at some
temperatures, the signature of orientational relaxation is
observed as a nonoscillatory decay component in the signal.

\begin{figure}
\includegraphics[scale=0.82]{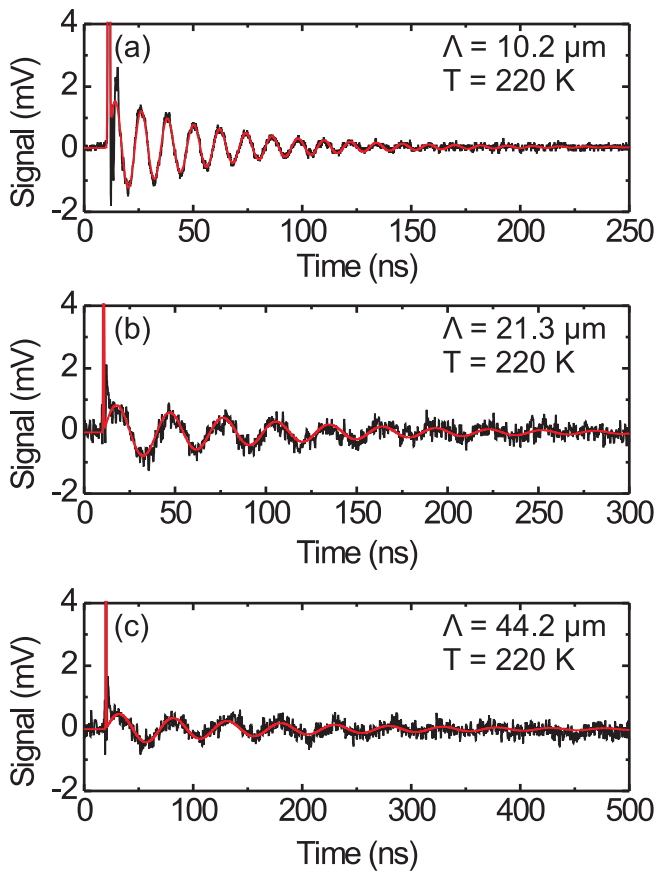}
\caption{\label{fig:TPP_220K}Shear acoustic signal in TPP at 220~K
for (a) 10.2~$\mu$m, (b) 21.3~$\mu$m, and (c) 44.2~$\mu$m grating
spacings. The data are in black, and the fits are in red. We note
that already by 220~K at 44.2~$\mu$m, only a few acoustic cycles
are observed, and that the signal intensity is reduced.}
\end{figure}

Another illustration of the influence of relaxation dynamics on
the signal may be obtained by examining data from a collection of
wavevectors at a common temperature, as depicted in
Fig.~\ref{fig:TPP_220K} where we provide data recorded with
10.2~$\mu$m, 21.3~$\mu$m, and 44.2~$\mu$m grating spacings at
220~K. Data with a fourth wavelength, 2.33~$\mu$m, are shown in
Fig.~\ref{fig:TPP_2.4}a. As the wavevector and the frequency are
reduced, the acoustic oscillation period increases toward the
characteristic relaxation timescale $\tau_s$ and therefore the
shear wave is more heavily damped. The signals at larger grating
spacings are weaker due to the linear $q$-dependence of the
excitation efficiency.

Based on the analysis of Sec.~\ref{sec:thry}, time-domain signals
were fit to the function
\begin{equation}\label{eq:6shearsig}
I(t)=A'\exp(-\Gamma_s t)\sin(\omega_s t+\phi)+B'\exp(-\Gamma_R t)
+ C\delta(t)
\end{equation}
which was convolved with the instrument response function,
modelled here by a Gaussian with duration 0.262~ns. The
convolution was necessary to determine the true $t=0$ for the
experiment. Here, $A'$ is the acoustic amplitude, $\Gamma_s$ is
the shear damping rate and $\omega_s$ is the shear frequency.
$\phi$ is a phase which accounts for the cosine term in
Eq.~\ref{eq:shearsol}, which only becomes important when the
damping is strong. In the next term, $B'$ is the optical Kerr
effect signal amplitude and $\Gamma_R$ is the orientational
relaxation rate, and in the last term $C$ is the strength of the
hyperpolarizability spike. As discussed in Sec.~\ref{sec:thry}, it
is a simplification to model the orientational behavior by a
single decaying exponential \cite{fayer}. A more accurate
description might be in terms of a Kohlrausch-Williams-Watts
stretched exponential function $e^{\left(-t/tau_s\right)^\beta}$
($0<\beta\leq 1$), as is used commonly for time-domain relaxation;
however, since the orientational signal contributions are weak, we
were able to obtain excellent fits with fewer parameters using a
single exponential form.

\begin{figure}
\centering
\includegraphics[scale=0.95]{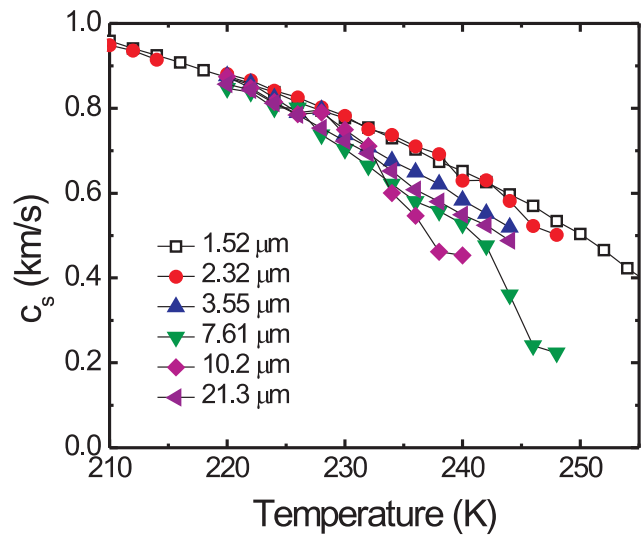}
\caption{\label{fig:tppcs}Shear speed of sound in TPP as a
function of temperature for a variety of grating spacings. We note
that the highest temperature for which we could observe shear
waves increases with the decrease of grating spacing. Data at
$\Lambda=~1.52~\mu$m are from \cite{sil1,sil2}.}
\end{figure}

\begin{figure}
\includegraphics[scale=0.95]{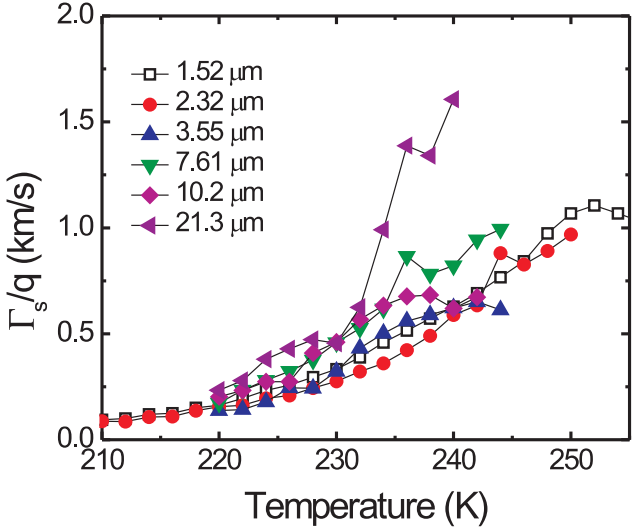}
\caption[$q \Gamma_s(T)$ in TPP for a variety of grating
spacings]{\label{fig:tppdamp}Scaled shear damping rate in TPP as a
function of temperature for a variety of grating spacings. Longer
wavelength acoustic waves have a higher scaled damping rate which
increases with temperature. Data at $\Lambda=~1.52~\mu$m are from
\cite{sil1,sil2}.}
\end{figure}

The obtained values of the the shear acoustic velocity
$c_s=\omega_s/q$ at a collection of wavevectors are shown in
Fig.~\ref{fig:tppcs}, while in Fig.~\ref{fig:tppdamp} the scaled
damping rates are shown. Both figures incorporate the data from
Refs.~\cite{sil1} and \cite{sil2}. Data at other wavevectors were
consistent with those shown, but are omitted from this and further
plots for clarity. Two features of the data are immediately
evident in these plots: first, we observe significant acoustic
dispersion for the shear waves across all temperatures studied,
and this dispersion increases dramatically as the temperature is
raised. The second feature we note is that, as a result of the
dispersion and the shear softening it represents, at each
temperature above 240~K there is a wavelength above which we are
unable to observe the shear wave in our measurements due to its
increased damping and reduced signal strength. This wavelength is
observed to decrease as temperature increases.

\begin{figure}
\includegraphics[scale=0.95]{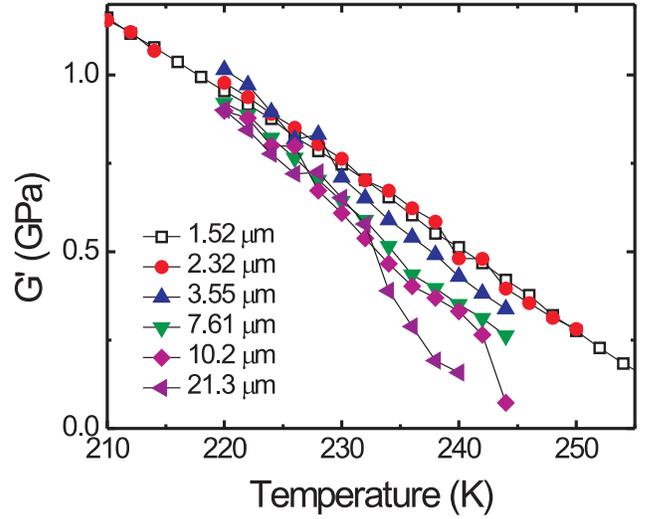}
\caption[$G'(T)$ at a number of grating
spacings]{\label{fig:tpprefogt}$G'(T)$ at a number of grating
spacings. As the temperature is increased, the shear modulus is
observed to soften considerably. Again, the highest temperature
for which we could observe shear waves increases with the decrease
of grating spacing. Data at $\Lambda=~1.52~\mu$m are from
\cite{sil1,sil2}.}
\end{figure}

\begin{figure}
\includegraphics[scale=0.95]{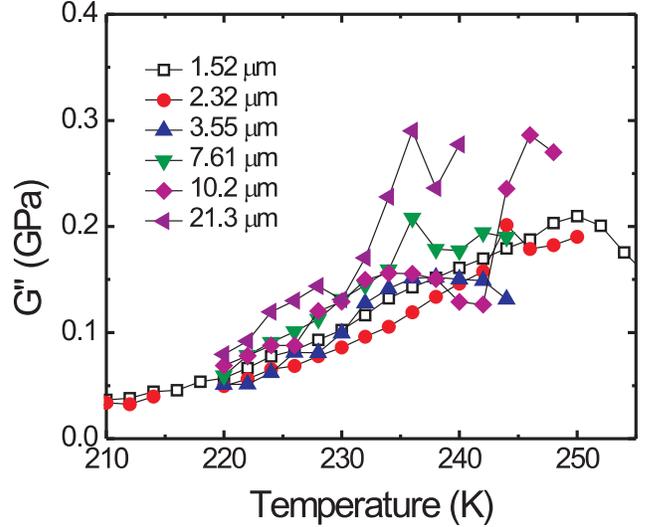}
\caption[$G''(T)$ at a number of grating
spacings]{\label{fig:tppimfogt}$G''(T)$ at a number of grating
spacings. For most wavevectors examined in this study, we only
observed a rise in the value of $G''(T)$ with temperature. At a
handful of wavevectors, we were able to see the low-frequency side
of the relaxation peak through, as is visible for
$2\pi/q=1.52~\mu$m and $10.2~\mu$m. Data at $\Lambda=1.52~\mu$m
are from \cite{sil1,sil2}.}
\end{figure}

From the fitted values for the shear frequency $\omega_s$ and the
damping rate $\Gamma_s$, we may compute a value of the reactive
and dissipative shear moduli from equations \ref{realmod} and
\ref{imagmod}. The density values we obtained from data in
reference \cite{TPPDEN} were fit to a quadratic function as
\begin{equation}\label{eq:tpp_den}
\rho (T)=1.507~[{\rm g/cm^3}]-1.3\times 10^{-3}~T~[{\rm K}]+6.8
\times 10^{-7}~T^2 ~[{\rm K^2}].
\end{equation}
Figures~\ref{fig:tpprefogt} and \ref{fig:tppimfogt} show plots of
the real and imaginary parts of the shear modulus, respectively,
as functions of temperature. These plots are for the same
collection of wavevectors for which we have plotted the velocity
and damping information. As in the plot of the velocity, we see
the softening of the modulus at higher temperatures. The imaginary
part shows generally monotonic behavior as a function of
temperature as well, except for the 1.52~$\mu$m and 10.2~$\mu$m
data, which show a small decrease in the imaginary part at higher
temperatures, a feature which is only weakly evident in the
damping rate itself represented in Fig.~\ref{fig:tppdamp}. This is
likely due to the oscillation period of our shear wave exceeding
the characteristic relaxation time $\tau_s$, permitting
observation of a piece of the low-frequency side of the relaxation
curve. We generally did not observe this trend at most
wavevectors, as the shear wave signal became either too weak or
too strongly damped to be observed.

The moduli at each temperature were plotted as a function of
frequency and then fit to the Havriliak-Negami relaxation function
\begin{equation}\label{eq:Ginf}
G^{\ast}(\omega_s)=G_{\infty}\left(1-\frac{1}{(1+(i\omega_s\tau_s)^\alpha)^\beta}\right)
\end{equation}
in order to extract the shear relaxation spectrum ($G_0=0$ for all
temperatures, by definition). Since time-temperature superposition
has been observed to hold for $\alpha$ relaxation in many liquids
and in TPP (at lower temperatures than measured here)
\cite{olsen_dyre}, all spectra were fit simultaneously with the
spectral parameters $\alpha$ and $\beta$ acting as
temperature-independent global variables and $\tau_s$. The value
for the temperature dependent infinite shear modulus
$G_{\infty}=\rho c_{\infty}^2$ was taken from depolarized
Brillouin scattering measurements performed by Chappell and
Kivelson~\cite{chappellkivelson} in a linear extrapolation to
colder temperatures as
\begin{equation}
c_{\infty}=2620.4~[{\rm m/s}] - 7.93 T~[{\rm K}]
\end{equation}
where the above expression has been obtained by using polarized
(VV) Brillouin scattering data from longitudinal acoustic phonons
to derive a temperature dependent refractive index~\cite{TPPDEN}
in combination with the shear data of the reference. We note that
these data fall on a common line with our lower-$T$ shear data of
Fig.~\ref{fig:tpprefogt} at the highest frequencies, i.e., the
shortest wavelengths ($\Lambda$ = 1.52~$\mu$m and 2.33~$\mu$m).

\begin{figure}
\centering
\includegraphics[scale=0.85]{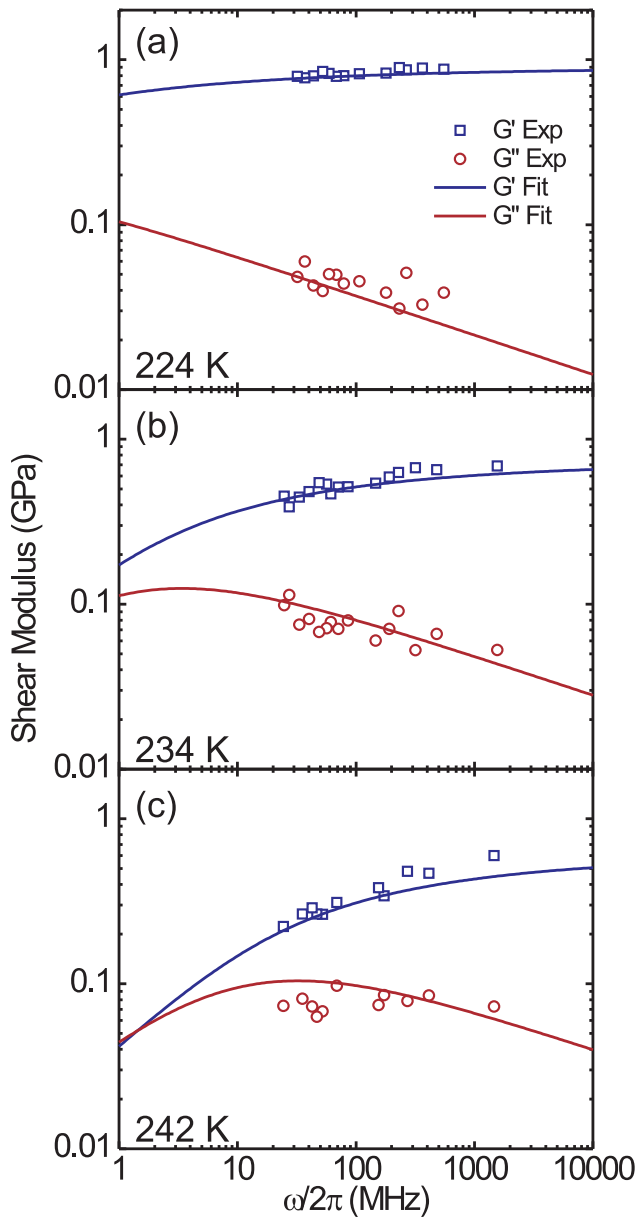}
\caption{\label{fig:TPP_GW}Plots of the real ($G'$) and imaginary
($G''$) moduli of TPP at (a) 224~K, (b) 234~K, and (c) 242~K. At
each temperature, a different segment of the shear acoustic
relaxation spectrum is present within the experimental frequency
window.}
\end{figure}

Three representative plots of the complex shear modulus with
corresponding fits are shown in Figs.~\ref{fig:TPP_GW}a -
\ref{fig:TPP_GW}c. As the temperature is increased, the shear
relaxation spectrum moves into the probed region. The fits
produced spectral parameters $\alpha=0.61$ and $\beta=0.39$.
Although not shown here, the longitudinal data from
Refs.~\cite{sil1,sil2} were refit using the same procedure for
purposes of consistency and comparison, yielding spectral
parameters $\alpha=0.69$ and $\beta=0.30$.

\begin{figure}
\includegraphics[scale=0.9]{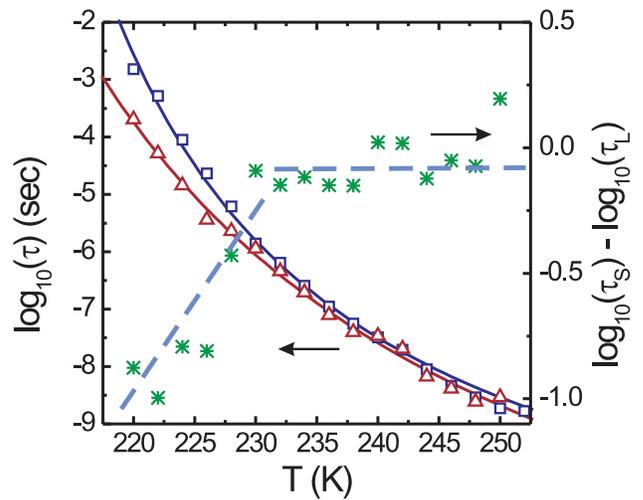}
\caption{\label{fig:vft_tpp}Characteristic relaxation time for
shear ($\tau_s$) and longitudinal ($\tau_l$) degrees of freedom
plotted versus temperature (left ordinate). Refitted values of the
VFT parameters of the original longitudinal data are
$\tau_0=2.5~\pm 1$~$\mu$s, $B=230~\pm 40$~K, and $T_0=198~\pm 4$~K
while the shear fits produced $\tau_0=0.26~\pm 0.2$~$\mu$s,
$B=450~\pm 240$~K, and $T_0=180~\pm 14$~K. Also shown is the
decoupling parameter $\log{\tau_s}-\log{\tau_l}$ as a function of
temperature (right ordinate).}
\end{figure}

Fig.~\ref{fig:vft_tpp} presents the fitted values of $\tau_s$ and
$\tau_l$ as a function of temperature along with their respective
VFT fits. There is excellent agreement between the two timescales
up to the temperature $T=231$~K, where the characteristic
relaxation times clearly separate. When combined with the
observation that the HN spectral parameters $\alpha$ and $\beta$
differ between the two degrees of freedom, we conclude that the
shear acoustic wave dynamics differ significantly, at least at the
lower sample temperatures measured, from those obtained from the
earlier polarized ISTS data \cite{sil1,sil2}. This conclusion,
based on a larger data set taken with a broader range of acoustic
wavevectors, supercedes the results of the previously published
work.

We now use our data to examine the phenomenon of $\alpha$-scale
coupling described briefly in the Introduction. Using the fitted
values of $\tau_s$ and $\tau_l$ as a function of temperature, we
define the coupling parameter as $\log{\tau_s}-\log{\tau_l}$,
which is plotted alongside the relaxation time in
Fig.~\ref{fig:vft_tpp}. The coupling parameter is constant and
essentially zero until it begins to decrease with decreasing
temperature, reflecting the differing characteristic timescales of
the shear and longitudinal relaxation as the former becomes
slower. This provides an estimate of the MCT crossover temperature
$T_c=231$.

\begin{figure}
\includegraphics[scale=1]{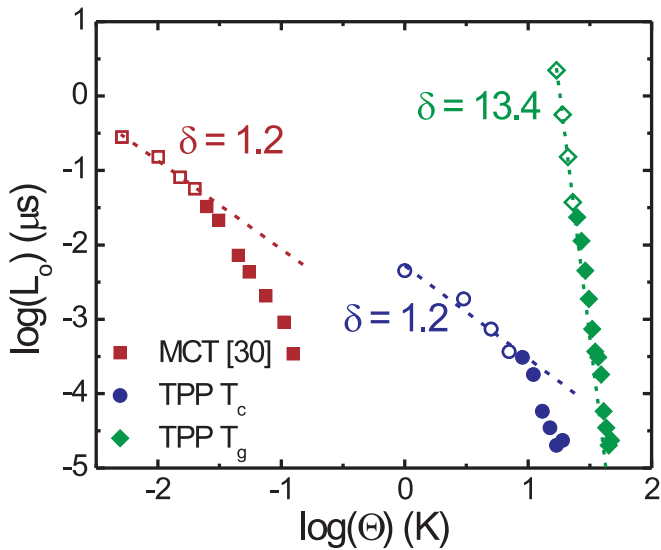}
\caption{\label{fig:length_tpp}Computed value of length scale
limit for shear wave propagation as a function of temperature
$\Theta = T-T'$ using the decoupling value for $T'=T_c$ (blue
circles) as the critical temperature and the glass transition
temperature $T'=T_g$ (green diamonds). For comparison, results
from Ref.~\cite{das1} are also shown as a function of packing
fraction $\Delta$ (red squares). The exponents are from the data
plotted with empty symbols.}
\end{figure}

We may attempt to further understand our results in the
mode-coupling framework of Ahluwalia and Das \cite{das1}. Briefly,
when a calculation of a collection of hard spheres in a
Percus-Yevick approximation is considered, the critical length
scale $L_0=2\pi/q_0$ above which propagation of shear waves
becomes overdamped obeys the power law in the vicinity of the
critical control parameter
\begin{equation}\label{eq:dasshear}
L_0=\frac{A}{(\Delta_c-\Delta)^{\delta}}
\end{equation}
where $A=1$, $\delta=1.2$, $\Delta$ represents the packing
fraction, and $\Delta_c$ is the MCT critical packing fraction
beyond which shear wave propagation is allowed for all length
scales.

Using the theoretical results of Sec.~\ref{sec:thry}, we may
attempt to deduce a lower length scale for shear wave propagation
as a function of temperature by considering at which wavevector
shear wave propagation becomes overdamped. For this analysis, we
chose the temperature $T$ to be the independent parameter,
yielding a similar power law $L_0=A/(T-T')^{\delta}$, where $T'$
represents a critical temperature. Here we consider two
significant temperatures for the liquid, i.e., the glass
transition temperature $T_g=202$~K \cite{TPPDI} and the crossover
temperature $T_c$ as determined from the MCT decoupling analysis
above.

Figure~\ref{fig:length_tpp} shows a plot of the derived upper
bound, as deduced from Eq.~\ref{eq:cicond} as a function of the
variable $\Theta = (T-T')$. The results of Ref.~\cite{das1} are
also shown as a function of packing fraction $\Delta$ for
comparison. A power law fit is shown to the four points in the
vicinity of the relevant temperatures. Picking $T'=T_c$ as the
relevant temperature yields a fitted value of $\delta=1.2$, which
is in excellent agreement with the theoretical result. A similar
fit to the data using the literature value of $T'=T_g$ produces a
significantly higher value of $\delta=13.4$. We remark that the
wavelength scales $L_0$ reached as this temperature is approached
from above are several orders of magnitude larger than those
corresponding to any diverging structural correlation length
scale.

\section{Conclusions}
We have used depolarized impulsive stimulated Brillouin scattering
to measure shear acoustic waves in supercooled triphenyl phosphite
from 220~K to 250~K combined with previously obtained results, we
are able to examine a frequency regime from $\sim 10$~MHz to
almost $1$~GHz. Our results indicate that the shear and
longitudinal spectra do not share the same spectral parameters
$\alpha$ and $\beta$.

We also observed $\alpha$-decoupling of the longitudinal and shear
degrees of freedom, yielding an estimate of the mode-coupling
$T_c=231$~K. Using the decoupling result, we verified a power law
for the diverging lengthscale of shear wave propagation as
$\delta=1.24$. A similar test using the literature value of $T_g$
was not in agreement with the theoretical model, as should be
expected since $T_g$ is not a mode-coupling theory parameter.

Further work in the study of shear relaxation in triphenyl
phosphite and other liquids will center on expanding the dynamic
range of the measurements. We also note that a comparison with
dielectric data via the model of DiMarzio and Bishop \cite{dim1}
could also be performed if dielectric relaxation measurements at a
similar combination of temperatures and frequencies were carried
out.

\section{Acknowledgments}

We gratefully acknowledge Professor Shankar P. Das for useful
discussions. This work was supported in part by National Science
Foundation Grants No. CHE-0616939 and IMR-0414895.

\bibliography{TPPShear}

\end{document}